\documentclass[12pt]{article}
\usepackage{amssymb,amsfonts}
\usepackage{epsf,epsfig}
\usepackage{graphicx}
\graphicspath{ {images/} }
\usepackage{caption}
\usepackage{subcaption}
\usepackage{amssymb}

\begin{document}
\baselineskip 18pt

\title{Thermodynamics of an Ising-like $XXZ$ chain in a longitudinal magnetic field in the framework of the Quantum Transfer Matrix approach}

\author{P.~N.~Bibikov}
\date{\it Russian State Hydrometeorological University, Saint-Petersburg, Russia}

\maketitle

\vskip5mm

\begin{abstract}
Taking the Ising chain as a reference model we have derived a perturbative expression for the free energy density of the Heisenberg-Ising chain with strong easy-axis anisotropy.
All calculations are performed on the ground of the Quantum Transfer Matrix approach. The obtained result agrees with the direct high-temperature expansion.
It also agrees with the low-temperature cluster expansion in the special subregime when quantum fluctuations are weak against thermodynamical ones.
\end{abstract}

\maketitle

\section{Introduction}

One of the basic models of one-dimensional quantum magnetism is the $XXZ$ spin chain in a {\it longitudinal} magnetic field $h$ \cite{1}. It corresponds to the Hamiltonian
\begin{equation}
\hat H=\sum_{n=1}^N\Big[\frac{J}{2}\Big({\bf S}^+_n{\bf S}^-_{n+1}+{\bf S}^-_n{\bf S}^+_{n+1}\Big)+
J_z\Big({\bf S}^z_n{\bf S}^z_{n+1}-\frac{1}{4}I\Big)-h{\bf S}_n^z\Big],\qquad h\geq0,
\end{equation}
where ${\bf S}_n^z$ and ${\bf S}_n^{\pm}={\bf S}_n^x\pm i{\bf S}_n^y$ are the usual spin-1/2 operators.
At $J=0$ the model (1) reduces to the Ising chain solvable by a rather simple machinery \cite{2}.

Being purely {\it classical} ($[{\bf S}_m^z,{\bf S}_n^z]=0$) the Ising model is rather poor (for example the Ising magnons are dispersionless). Nevertheless it was suggested for a number
of magnetic compounds \cite{3,4,5}.
However since the condition $J=0$ does not follow from any symmetry it is natural to suppose that a more adequate model for these compounds is the Ising-like chain described by Hamiltonian (1) 
supplemented by the condition
\begin{equation}
\Delta\equiv\frac{J_z}{J}\gg1.
\end{equation}

The model (1), (2) was also suggested for some real compounds \cite{6,7}. Being {\it quantum} it has a more rich physical behavior and at the same time should allow a perturbative treatment around the Ising model.

In the last two decades a new machinery for evaluation of thermodynamics of quantum spin chains was suggested basing on the Quantum Transfer Matrix (QTM) approach
(see reviews \cite{8,9} and references therein). The latter has two main stages.
Within the former one a system of integral equations on specially introduced auxiliary functions is derived. Within the latter one an integral representations for the free energy density
\begin{equation}
f(T,h)=-\frac{1}{\beta}\lim_{N\rightarrow\infty}\frac{1}{N}\ln{Z_N(T,h)},
\end{equation}
(as usual $Z_N(T,h)$ is the partition function and $\beta\equiv1/T$) and correlation functions are obtained in this framework.

In the Ising $J=0$ case the corresponding free energy density has the form
\begin{equation}
f^{Is}(T,h)=e_0(h)-\frac{1}{\beta}\ln{\left(\frac{1+{\rm e}^{-\beta h}+\sqrt{(1-{\rm e}^{-\beta h})^2+4{\rm e}^{\beta(J_z-h)}}}{2}\right)},
\end{equation}
where
\begin{equation}
e_0(h)=-\frac{h}{2}
\end{equation}
is the ground state energy density. Formula (4) readily follows from the representation \cite{2}
\begin{equation}
f^{Is}(T,h)=-\frac{1}{\beta}\ln{\Lambda_+(\beta,h)},
\end{equation}
where
\begin{equation}
\Lambda_{\pm}(\beta,h)=\cosh{\frac{\beta h}{2}}\pm\sqrt{\sinh^2{\frac{\beta h}{2}}+{\rm e}^{\beta J_z}},
\end{equation}
are eigenvalues of the special transfer matrix
\begin{equation}
T=\left(\begin{array}{cc}
{\rm e}^{\beta h/2}&{\rm e}^{\beta J/2}\\
{\rm e}^{\beta J/2}&{\rm e}^{-\beta h/2}
\end{array}\right),
\end{equation}
related to the Hamiltonian (1) at $J=0$. As it has been mentioned in \cite{10} formula (4) also may be obtained within the approach of \cite{8,9}.

In the present paper we extend the result of \cite{10} studying the model (1), (2) at the vicinity of the Ising point in the first two perturbation orders. The paper is organized as follows.
In Sect. 2 following \cite{10} we introduce the basic auxiliary functions and corresponding integral equations. We use however rather different notations which seem us to be more convenient
for our treatment. We also show how to account singularities in the kernels of the integral equations. In Sect. 3 we extract
the auxiliary functions related to the Ising model \cite{10} and reduce the integral equations to the form convenient for perturbative expansion.
In Sect. 4 we calculate the first two terms of the perturbation expansion and evaluate the corresponding correction to the free energy density.
In Sect. 5 treating the high-temperature regime
\begin{equation}
h,|J_z|,|J|\ll k_BT\Longleftrightarrow\beta\cdot\max{(h,|J_z|,|J|)}\ll1,
\end{equation}
we compare the first order terms of the high-temperature expansion which follows from the obtained formula for $f(T,h)$ with the one directly related to (3).
Showing that both approaches give the same result we confirm the effectiveness of the approach \cite{8,9} at high temperatures \cite{11}.
In Sect. 6 we study the low-temperature regime in the phase related to the ferromagnetically polarized ground state
\begin{equation}
|\emptyset\rangle=\dots\otimes|\uparrow\rangle\otimes|\uparrow\rangle\otimes|\uparrow\rangle\otimes\dots,\qquad{\bf S}^z|\uparrow\rangle=\frac{1}{2}|\uparrow\rangle,
\end{equation}
and non-zero magnon gap energy \cite{12}
\begin{equation}
E_{gap}=h-J_z-|J|>0.
\end{equation}
Comparing the calculated peturbative result with the one obtained previously by the low-temperature cluster expansion \cite{12} we show that in the {\it extreme}
low-temperature subregime
\begin{equation}
k_BT\ll\frac{E_{width}}{2},
\end{equation}
where
\begin{equation}
E_{width}=2|J|,
\end{equation}
is the magnon band width, the both approaches totally disagree. However they give similar results in the {\it scaled} low-temperature subregime
\begin{equation}
\frac{E_{width}}{2}\ll k_BT\ll E_{gap}+\frac{E_{width}}{2}.
\end{equation}

If one suggest that low-temperature quantum fluctuations arise from transitions within the magnon band then the associated fluctuation of energy should
be $E_{width}$. Hence the scaled low-temperature regime corresponds to weakness of quantum fluctuations against thermodynamical ones.

\section{Foundations of the ${\mathfrak b}_+$-${\mathfrak b}_-$-formalism}

Following \cite{10} (however with unconventional notations $\mathfrak b_+(x)$ and $\mathfrak b_-(x)$ instead of
$\mathfrak b(x)$ and $\bar\mathfrak b(x)$) we suggest the following system of equations
\begin{equation}
\ln{\mathfrak b_{\pm}}(x)+\int_{-\pi/2}^{\pi/2}dy\Big(\kappa_{\mp}(x-y)\ln{\mathfrak B_{\mp}}(y)-\kappa(x-y)
\ln{\mathfrak B_{\pm}}(y)\Big)=\mp\frac{\beta h}{2}-J_z\beta c(x),
\end{equation}
where
\begin{equation}
\mathfrak B_{\pm}(x)=1+\mathfrak b_{\pm}(x),
\end{equation}
and the $T=\infty$ condition
\begin{equation}
\lim_{T\rightarrow\infty}{\mathfrak b_{\pm}}(x,T,h)=1.
\end{equation}
Here
\begin{eqnarray}
&&d(x)=\sum_{m=-\infty}^{\infty}\frac{{\rm e}^{2imx}}{{\rm ch}{(\eta m)}},\qquad c(x)=\frac{d(x)\tanh{\eta}}{2},\qquad
\kappa(x)=\frac{1}{\pi}\sum_{m=-\infty}^{\infty}\frac{{\rm e}^{2imx}}{1+{\rm e}^{2\eta|m|}},\nonumber\\
&&\kappa_{\pm}(x)=\kappa(x\pm i\eta\mp i\epsilon)=\frac{1}{\pi}\sum_{m=-\infty}^{\infty}\frac{{\rm e}^{\mp2(\eta-\epsilon)m}}{1+{\rm e}^{2\eta|m|}}{\rm e}^{2imx},
\end{eqnarray}
and it is assumed that
\begin{equation}
\cosh{\eta}=\frac{J_z}{J},\qquad\eta>0.
\end{equation}

Functions $\kappa_{\pm}(x)$ have singular parts. In order to extract them we shall use the following extended representations
\begin{eqnarray}
&&\kappa_+(x)=\frac{1}{\pi}\Big[\frac{1}{2}+\sum_{m=1}^{\infty}\frac{{\rm e}^{2(ix-\eta)m}}{1+{\rm e}^{2\eta m}}+\sum_{m=-\infty}^{-1}\Big(1-\frac{1}{1+{\rm e}^{2\eta|m|}}\Big){\rm e}^{2(ix+\epsilon)m}\Big],\nonumber\\
&&\kappa_-(x)=\frac{1}{\pi}\Big[\frac{1}{2}+\sum_{m=1}^{\infty}\Big(1-\frac{1}{1+{\rm e}^{2\eta m}}\Big){\rm e}^{2(ix-\epsilon)m}+\sum_{m=-\infty}^{-1}\frac{{\rm e}^{2(ix+\eta)m}}{1+{\rm e}^{2\eta|m|}}\Big].
\end{eqnarray}
Hence for a function
\begin{equation}
\varphi(x)=\varphi^{(0)}+\varphi^{(+)}(x)+\varphi^{(-)}(x),
\end{equation}
where $\varphi^{(0)}$ is a number and
\begin{equation}
\varphi^{(+)}(x)=\sum_{m=1}^{\infty}\varphi^{(m)}{\rm e}^{2imx},\qquad\varphi^{(-)}(x)=\sum_{m=1}^{\infty}\varphi^{(-m)}{\rm e}^{-2imx},
\end{equation}
one has
\begin{equation}
\int_{-\pi/2}^{\pi/2}\kappa_{\pm}(x-y)\varphi(y)dy=\frac{\varphi^{(0)}}{2}+\varphi^{(\mp)}(x)+\int_{-\pi/2}^{\pi/2}\tilde\kappa_{\pm}(x-y)(\varphi^{(+)}(y)+\varphi^{(+)}(y))dy,
\end{equation}
where
\begin{eqnarray}
&&\tilde\kappa_+(x)=\frac{1}{\pi}\left(\sum_{m=1}^{\infty}\frac{{\rm e}^{2(ix-\eta)m}}{1+{\rm e}^{2\eta m}}-\sum_{m=-\infty}^{-1}\frac{{\rm e}^{2ixm}}{1+{\rm e}^{2\eta|m|}}\right),\nonumber\\
&&\tilde\kappa_-(x)=\frac{1}{\pi}\left(-\sum_{m=1}^{\infty}\frac{{\rm e}^{2ixm}}{1+{\rm e}^{2\eta m}}+\sum_{m=-\infty}^{-1}\frac{{\rm e}^{2(ix+\eta)m}}{1+{\rm e}^{2\eta|m|}}\right).
\end{eqnarray}

With the use of the auxiliary functions ${\mathfrak B}_{\pm}(x)$ the free energy density (3) may be represented by the formula \cite{10}
\begin{equation}
f(T,h)=-\frac{J_z\tanh{\eta}}{2}\sum_{m=-\infty}^{\infty}\frac{{\rm e}^{-\eta|m|}}{\cosh{(\eta m)}}-\frac{1}{2\pi\beta}\int_{-\pi/2}^{\pi/2}dxd(x)\ln{({\mathfrak B}_+(x){\mathfrak B}_-(x))}.
\end{equation}

\section{The Ising solution and around}

At the Ising point
\begin{equation}
J=0\Longleftrightarrow\eta=\infty,
\end{equation}
one has from (18) and (24)
\begin{equation}
d^{Is}(x)=1,\qquad c^{Is}(x)=\frac{1}{2},\qquad\kappa^{Is}(x)=\frac{1}{2\pi},\qquad\tilde\kappa^{Is}_{\pm}(x)=0,
\end{equation}
so the right side of (15) does not depend on $x$ and a substitution
\begin{equation}
{\mathfrak b}_{\pm}(x)\longrightarrow{\mathfrak b}_{\pm}^{Is},\qquad{\mathfrak B}_{\pm}(x)\longrightarrow{\mathfrak B}_{\pm}^{Is}=1+{\mathfrak b}_{\pm}^{Is}
\end{equation}
yields the following system of algebraic equations
\begin{equation}
\ln{\mathfrak b_{\pm}^{Is}}+\frac{\ln{\mathfrak B_{\mp}^{Is}}-\ln{\mathfrak B_{\pm}^{Is}}}{2}=\mp\frac{\beta h}{2}-\frac{J_z\beta}{2}\Longleftrightarrow
({\mathfrak b}_{\pm}^{Is})^2\frac{1+{\mathfrak b}_{\mp}^{Is}}{1+{\mathfrak b}_{\pm}^{Is}}={\rm e}^{\beta(-J_z\mp h)},
\end{equation}
or in an equivalent form
\begin{equation}
{\mathfrak b}_+^{Is}{\mathfrak b}_-^{Is}={\rm e}^{-\beta J_z},\qquad
\frac{({\mathfrak b}_{\pm}^{Is})^2+{\rm e}^{-\beta J_z}{\mathfrak b}_{\pm}^{Is}}{{\mathfrak b}_{\pm}^{Is}+1}={\rm e}^{\beta(-J_z\mp h)}.
\end{equation}
At the same time formula (25) in the Ising case (26) reduces to
\begin{equation}
f^{Is}(T,h)=-\frac{J_z}{2}-\frac{1}{2\beta}\ln{({\mathfrak B}^{Is}_+{\mathfrak B}^{Is}_-)}.
\end{equation}

System (30), (17) has the single solution
\begin{equation}
{\mathfrak b}_{\pm}^{Is}={\rm e}^{-\beta(J_z\pm h/2)}\Big(\sqrt{\sinh^2{\frac{\beta h}{2}}+{\rm e}^{\beta J_z}}\mp\sinh{\frac{\beta h}{2}}\Big).
\end{equation}
According to (16) and (30)
\begin{equation}
{\mathfrak B}_+^{Is}{\mathfrak B}_-^{Is}=1+{\rm e}^{-\beta J_z}+{\mathfrak b}_+^{Is}+{\mathfrak b}_-^{Is},
\end{equation}
so a substitution of (32) into (33) gives
\begin{eqnarray}
&&{\mathfrak B}_+^{Is}{\mathfrak B}_-^{Is}=1+{\rm e}^{-\beta J_z}\Big(2\cosh{\frac{\beta h}{2}}\cdot\sqrt{\sinh^2{\frac{\beta h}{2}}+{\rm e}^{\beta J_z}}+\cosh^2{\frac{\beta h}{2}}
+\sinh^2{\frac{\beta h}{2}}\Big)\nonumber\\
&&={\rm e}^{-\beta J_z}\Big(\cosh{\frac{\beta h}{2}}+\sqrt{\sinh^2{\frac{\beta h}{2}}+{\rm e}^{\beta J_z}}\Big)^2.
\end{eqnarray}
So (4) really follows from (31) and (34) \cite{10}.

Let us now extract the Ising solution taking
\begin{equation}
{\mathfrak b}_{\pm}(x)={\mathfrak b}_{\pm}^{Is}+{\tilde\mathfrak b}_{\pm}(x),\qquad f(T,h)=f^{Is}(T,h)+\tilde f(T,h),
\end{equation}
as well as
\begin{equation}
d(x)=d^{Is}+\tilde d(x),\qquad c(x)=c^{Is}+\tilde c(x),\qquad\kappa(x)=\kappa^{Is}+\tilde\kappa(x),
\end{equation}
where according to (18), (27) and (36)
\begin{eqnarray}
&&\tilde d(x)=2\sum_{m=1}^{\infty}\frac{\cos{2mx}}{\cosh{\eta m}},\qquad\tilde c(x)=\tanh\eta\sum_{m=1}^{\infty}\frac{\cos{2mx}}{{\rm ch}{(\eta m)}}-\frac{{\rm e}^{-2\eta}}{1+{\rm e}^{-2\eta}},\nonumber\\
&&\tilde\kappa(x)=\frac{2}{\pi}\sum_{m=1}^{\infty}\frac{\cos{2mx}}{1+{\rm e}^{2\eta m}}.
\end{eqnarray}
From (24) and (37) readily follows that
\begin{equation}
\int_{-\pi/2}^{\pi/2}\tilde d(x)dx=\int_{-\pi/2}^{\pi/2}\tilde\kappa(x)dx=\int_{-\pi/2}^{\pi/2}\tilde\kappa_{\pm}(x)dx=0.
\end{equation}
Using now (35)-(38) one may reduce (15) and (25) to the forms
\begin{eqnarray}
&&\ln{\Big(1+\frac{\tilde\mathfrak b_{\pm}(x)}{\mathfrak b_{\pm}^{Is}}\Big)}+\int_{-\pi/2}^{\pi/2}dy\Big[
\kappa_{\mp}(x-y)\ln{\Big(1+\frac{\tilde\mathfrak b_{\mp}(y)}{\mathfrak B_{\mp}^{Is}}\Big)}
\nonumber\\
&&-\kappa(x-y)\ln{\Big(1+\frac{\tilde\mathfrak b_{\pm}(y)}{\mathfrak B_{\pm}^{Is}}\Big)}\Big]=-J_z\beta\tilde c(x),
\end{eqnarray}
and
\begin{eqnarray}
&&\tilde f(T,h)=J_z\Big(\frac{1-\tanh{\eta}}{2}-2\tanh{\eta}\sum_{m=1}^{\infty}\frac{{\rm e}^{-2\eta m}}{1+{\rm e}^{-2\eta m}}\Big)\nonumber\\
&&-\frac{1}{2\pi\beta}\int_{-\pi/2}^{\pi/2}dxd(x)\Big[\ln{\Big(1+\frac{\tilde\mathfrak b_+(x)}{\mathfrak B_+^{Is}}\Big)}
+\ln{\Big(1+\frac{\tilde\mathfrak b_-(x)}{\mathfrak B_-^{Is}}\Big)}\Big],
\end{eqnarray}
more convenient for the perturbative series expansion.

\section{Series expansion near the Ising point}

Suggesting the series expansions
\begin{eqnarray}
&&{\tilde\mathfrak b}_{\pm}(x)=\sum_{j=1}^{\infty}{\rm e}^{-\eta j}{\mathfrak b}_{\pm}^{(j)}(x),\\
&&\tilde f(T,h)=\sum_{j=1}^{\infty}{\rm e}^{-\eta j}f^{(j)}(T,h),
\end{eqnarray}
and representing each ${\mathfrak b}_{\pm}^{(j)}(x)$ in the separated form (21)
\begin{equation}
{\mathfrak b}_{\pm}^{(j)}(x)={\mathfrak b}_{\pm}^{(j;0)}+{\mathfrak b}_{\pm}^{(j;+)}(x)+{\mathfrak b}_{\pm}^{(j;-)}(x),
\end{equation}
where
\begin{equation}
{\mathfrak b}_{\pm}^{(j;+)}(x)=\sum_{m=1}^{\infty}{\mathfrak b}_{\pm}^{(j;2m)}{\rm e}^{2imx},\qquad{\mathfrak b}_{\pm}^{(j;-)}(x)=\sum_{m=1}^{\infty}{\mathfrak b}_{\pm}^{(j;-2m)}{\rm e}^{-2imx},
\end{equation}
we shall calculate ${\mathfrak b}_{\pm}^{(j)}(x)$ and $f^{(j)}(T,h)$ for $j=1,2$. Since
\begin{equation}
\frac{1}{\Delta}\equiv\frac{1}{\cosh{\eta}}=2{\rm e}^{-\eta}+o({\rm e}^{-2\eta}),
\end{equation}
formula
\begin{equation}
\tilde f(T,h)={\rm e}^{-\eta}f^{(1)}(T,h)+{\rm e}^{-2\eta}f^{(2)}(T,h)+o({\rm e}^{-2\eta}),
\end{equation}
is equivalent to
\begin{equation}
\tilde f(T,h)=\frac{f^{(1)}(T,h)}{2\Delta}+\frac{f^{(2)}(T,h)}{4\Delta^2}+o\Big(\frac{1}{\Delta^2}\Big).
\end{equation}

For evaluation of the first two terms in the right sides of (41) and (42) we also need the expansions
\begin{eqnarray}
&&\tilde c(x)=2{\rm e}^{-\eta}\cos{2x}+(2\cos{4x}-1){\rm e}^{-2\eta}+o({\rm e}^{-2\eta}),\nonumber\\
&&\tilde\kappa(x)=\frac{2{\rm e}^{-2\eta}}{\pi}\cos{2x}+o({\rm e}^{-2\eta}),\nonumber\\
&&\tilde\kappa_{\pm}(x)=-\frac{{\rm e}^{2(\mp ix-\eta)}}{\pi}+o({\rm e}^{-2\eta}),\nonumber\\
&&d(x)=1+4({\rm e}^{-\eta}\cos{2x}+{\rm e}^{-2\eta}\cos{4x})+o({\rm e}^{-2\eta}),\nonumber\\
&&\frac{1-\tanh{\eta}}{2}-2\tanh{\eta}\sum_{m=1}^{\infty}\frac{{\rm e}^{-2\eta m}}{1+{\rm e}^{-2\eta m}}=-{\rm e}^{-2\eta}+o({\rm e}^{-2\eta}),
\end{eqnarray}
which directly follow from (37) and (24).

Using (48) one readily gets in the order ${\rm e}^{-\eta}$
\begin{equation}
f^{(1)}(T,h)=-\frac{1}{2\pi\beta}\int_{-\pi/2}^{\pi/2}\Big(\frac{\mathfrak b_+^{(1)}(x)}{\mathfrak B_+^{Is}}
+\frac{\mathfrak b_-^{(1)}(x)}{\mathfrak B_-^{Is}}\Big)dx=-\frac{1}{2\beta}\Big(\frac{\mathfrak b_+^{(1;0)}}{\mathfrak B_+^{Is}}
+\frac{\mathfrak b_-^{(1;0)}}{\mathfrak B_-^{Is}}\Big),
\end{equation}
and
\begin{equation}
\frac{\mathfrak b_{\pm}^{(1)}(x)}{\mathfrak b_{\pm}^{Is}}+\frac{\mathfrak b_{\mp}^{(1;\pm)}(x)}{\mathfrak B_{\mp}^{Is}}
+\frac{1}{2\pi}\int_{-\pi/2}^{\pi/2}\Big[\frac{\mathfrak b_{\mp}^{(1)}(y)}{\mathfrak B_{\mp}^{Is}}-\frac{\mathfrak b_{\pm}^{(1)}(y)}{\mathfrak B_{\pm}^{Is}}\Big]dy
=-2J_z\beta\cos{2x},
\end{equation}
or in an expanded form
\begin{eqnarray}
&&\frac{\mathfrak b_{\pm}^{(1;0)}}{\mathfrak b_{\pm}^{Is}}+\frac{1}{2}
\Big[\frac{\mathfrak b_{\mp}^{(1;0)}}{\mathfrak B_{\mp}^{Is}}-\frac{\mathfrak b_{\pm}^{(1;0)}}{\mathfrak B_{\pm}^{Is}}\Big]=0,\\
&&\frac{\mathfrak b_+^{(1;2)}(x)}{\mathfrak b_+^{Is}}+\frac{\mathfrak b_-^{(1;2)}(x)}{\mathfrak B_-^{Is}}=-J_z\beta,\qquad
\frac{\mathfrak b_+^{(1;-2)}(x)}{\mathfrak b_+^{Is}}=-J_z\beta,\nonumber\\
&&\frac{\mathfrak b_-^{(1;2)}(x)}{\mathfrak b_-^{Is}}=-J_z\beta,\qquad
\frac{\mathfrak b_-^{(1;-2)}(x)}{\mathfrak b_-^{Is}}+\frac{\mathfrak b_+^{(1;-2)}(x)}{\mathfrak B_+^{Is}}=-J_z\beta.
\end{eqnarray}

Representing homogeneous linear system (51) as
\begin{equation}
(2+\mathfrak b_{\pm}^{Is})t_{\pm}+\mathfrak b_{\pm}^{Is}t_{\mp}=0,\qquad t_{\pm}\equiv\frac{\mathfrak b_{\pm}^{Is}}{1+\mathfrak b_{\pm}^{Is}},
\end{equation}
one readily concludes that its nontrivial solvability implies condition $\mathfrak b_+^{Is}+\mathfrak b_-^{Is}=-2$ which can not be fulfilled
because both $\mathfrak b_{\pm}^{Is}$ in (32) are positive. Hence (51) yields
\begin{equation}
\mathfrak b_{\pm}^{(1;0)}=0,
\end{equation}
and according to (49) and (54)
\begin{equation}
f^{(1)}(T,h)=0.
\end{equation}

The remaining system (52) gives
\begin{eqnarray}
&&\mathfrak b_+^{(1;2)}=-\frac{J_z\beta\mathfrak b_+^{Is}}{\mathfrak B_-^{Is}},\qquad\mathfrak b_+^{(1;-2)}=-J_z\beta\mathfrak b_+^{Is}\nonumber\\
&&\mathfrak b_-^{(1;2)}=-J_z\beta\mathfrak b_-^{Is},\qquad\mathfrak b_-^{(1;-2)}=-\frac{J_z\beta\mathfrak b_-^{Is}}{\mathfrak B_+^{Is}}.
\end{eqnarray}

Turning to $f^{(2)}(T,h)$ and accounting (48) one readily gets from (40)
\begin{eqnarray}
&&f^{(2)}(T,h)=-J_z+\frac{1}{2\pi\beta}\int_{-\pi/2}^{\pi/2}\Big[-4\Big(\frac{\mathfrak b_+^{(1)}(x)}{\mathfrak B_+^{Is}}
+\frac{\mathfrak b_-^{(1)}(x)}{\mathfrak B_-^{Is}}\Big)\cos{2x}\nonumber\\
&&+\frac{1}{2}\Big(\frac{\mathfrak b_+^{(1)}(x)}{\mathfrak B_+^{Is}}\Big)^2-\frac{\mathfrak b_+^{(2)}(x)}{\mathfrak B_+^{Is}}
+\frac{1}{2}\Big(\frac{\mathfrak b_-^{(1)}(x)}{\mathfrak B_-^{Is}}\Big)^2-\frac{\mathfrak b_-^{(2)}(x)}{\mathfrak B_-^{Is}}\Big]dx\nonumber\\
&&=-J_z+\frac{1}{2\beta}\Big[-2\Big(\frac{\mathfrak b_+^{(1;2)}(x)}{\mathfrak B_+^{Is}}+\frac{\mathfrak b_-^{(1;2)}(x)}{\mathfrak B_-^{Is}}
+\frac{\mathfrak b_+^{(1;-2)}(x)}{\mathfrak B_+^{Is}}+\frac{\mathfrak b_-^{(1;-2)}(x)}{\mathfrak B_-^{Is}}\Big)\nonumber\\
&&-\frac{\mathfrak b_+^{(2;0)}}{\mathfrak B_+^{Is}}-\frac{\mathfrak b_-^{(2;0)}}{\mathfrak B_-^{Is}}
+\frac{\mathfrak b_+^{(1;2)}(x)\mathfrak b_+^{(1;-2)}(x)}{(\mathfrak B_+^{Is})^2}
+\frac{\mathfrak b_-^{(1;2)}(x)\mathfrak b_-^{(1;-2)}(x)}{(\mathfrak B_-^{Is})^2}\Big].
\end{eqnarray}
So for an evaluation of $f^{(2)}(T,h)$ we additionally need only $\mathfrak b_{\pm}^{(2;0)}(x)$. The latter may be extracted from the equation
\begin{eqnarray}
&&\frac{\mathfrak b_{\pm}^{(2)}(x)}{\mathfrak b_{\pm}^{Is}}-\frac{1}{2}\Big(\frac{\mathfrak b_{\pm}^{(1)}(x)}{\mathfrak b_{\pm}^{Is}}\Big)^2
+\frac{\mathfrak b_{\mp}^{(2;\pm)}(x)}{\mathfrak B_{\mp}^{Is}}-\frac{1}{2}\Big(\frac{\mathfrak b_{\mp}^{(1;\pm)}(x)}{\mathfrak B_{\mp}^{Is}}\Big)^2
+\frac{1}{2\pi}\int_{-\pi/2}^{\pi/2}\Big[\frac{\mathfrak b_{\mp}^{(2)}(y)}{\mathfrak B_{\mp}^{Is}}\nonumber\\
&&-\frac{1}{2}\Big(\frac{\mathfrak b_{\mp}^{(1)}(y)}{\mathfrak B_{\mp}^{Is}}\Big)^2
-\frac{\mathfrak b_{\pm}^{(2)}(y)}{\mathfrak B_{\pm}^{Is}}+
\frac{1}{2}\Big(\frac{\mathfrak b_{\pm}^{(1)}(y)}{\mathfrak B_{\pm}^{Is}}\Big)^2\Big]dy=J_z\beta(1-2\cos{4x}),
\end{eqnarray}
which follows from (39) in the order ${\rm e}^{-2\eta}$. According to (58)
\begin{equation}
\frac{\mathfrak b_{\pm}^{(2;0)}}{\mathfrak b_{\pm}^{Is}}
-\frac{\mathfrak b_{\pm}^{(1;2)}\mathfrak b_{\pm}^{(1;-2)}}{(\mathfrak b_{\pm}^{Is})^2}
+\frac{1}{2}\Big[\frac{\mathfrak b_{\mp}^{(2;0)}}{\mathfrak B_{\mp}^{Is}}
-\frac{\mathfrak b_{\mp}^{(1;2)}\mathfrak b_{\mp}^{(1;-2)}}{(\mathfrak B_{\mp}^{Is})^2}-\frac{\mathfrak b_{\pm}^{(2;0)}}{\mathfrak B_{\pm}^{Is}}
+\frac{\mathfrak b_{\pm}^{(1;2)}\mathfrak b_{\pm}^{(1;-2)}}{(\mathfrak B_{\pm}^{Is})^2}\Big]=J_z\beta.
\end{equation}

Taking now
\begin{eqnarray}
&&\mathfrak b_{\pm}^{(2;0)}\equiv\mathfrak B_{\pm}^{(Is)}x_{\pm},\\
&&A_{\pm}\equiv2J_z\beta+J_z^2\beta^2\Big(\frac{2}{\mathfrak B_{\mp}^{(Is)}}+\frac{(\mathfrak b_{\mp}^{(Is)})^2\mathfrak B_{\pm}^{(Is)}-(\mathfrak b_{\pm}^{(Is)})^2\mathfrak B_{\mp}^{(Is)}}{(\mathfrak B_{\pm}^{(Is)}\mathfrak B_{\mp}^{(Is)})^2}\Big),
\end{eqnarray}
and a substituting $\mathfrak b_{\pm}^{(1;2)}$ and $\mathfrak b_{\pm}^{(1;-2)}$ from (56) one reduces (59) to the form
\begin{equation}
\Big(\frac{2}{\mathfrak b_{\pm}^{(Is)}}+1\Big)x_{\pm}+x_{\mp}=A_{\pm},
\end{equation}
which yields
\begin{equation}
x_{\pm}=\frac{\mathfrak b_+^{(Is)}\mathfrak b_-^{(Is)}}{\mathfrak B_+^{(Is)}+\mathfrak B_-^{(Is)}}\Big(\frac{A_{\pm}}{\mathfrak b_{\mp}^{(Is)}}+\frac{A_{\pm}-A_{\mp}}{2}\Big).
\end{equation}
So according to (60) and (63)
\begin{equation}
\frac{\mathfrak b_+^{(2;0)}}{\mathfrak B_+^{Is}}+\frac{\mathfrak b_-^{(2;0)}}{\mathfrak B_-^{Is}}=x_++x_-=\frac{A_+\mathfrak b_+^{Is}+A_-\mathfrak b_-^{Is}}{\mathfrak B_+^{Is}+\mathfrak B_-^{Is}},
\end{equation}
and a direct substitution of (61) into (64) gives
\begin{equation}
A_+\mathfrak b_+^{Is}+A_-\mathfrak b_-^{Is}=2J_z\beta(\mathfrak U-2)+J_z^2\beta^2\Big[2\Big(\frac{\mathfrak U(\mathfrak U-1)}{\mathfrak V}-2\Big)+\frac{(1-\mathfrak V)(\mathfrak U^2-4\mathfrak V)}{\mathfrak V^2}\Big],
\end{equation}
where
\begin{equation}
{\mathfrak U}=\mathfrak B_+^{(Is)}+\mathfrak B_-^{(Is)},\qquad{\mathfrak V}=\mathfrak B_+^{(Is)}\mathfrak B_-^{(Is)}.
\end{equation}
Now from (64) and (65) follows that
\begin{equation}
\frac{\mathfrak b_+^{(2;0)}}{\mathfrak B_+^{Is}}+\frac{\mathfrak b_-^{(2;0)}}{\mathfrak B_-^{Is}}=
2J_z\beta\Big(1-\frac{2}{\mathfrak U}\Big)+J_z^2\beta^2\Big(\frac{\mathfrak U-2}{\mathfrak V}-\frac{4}{\mathfrak U\mathfrak V}+\frac{\mathfrak U}{\mathfrak V^2}\Big).
\end{equation}

At the same time according to (56)
\begin{eqnarray}
&&-\frac{\mathfrak b_+^{(1;2)}}{\mathfrak B_+^{Is}}-\frac{\mathfrak b_-^{(1;2)}}{\mathfrak B_-^{Is}}
-\frac{\mathfrak b_+^{(1;-2)}}{\mathfrak B_+^{Is}}-\frac{\mathfrak b_-^{(1;-2)}}{\mathfrak B_-^{Is}}
=2J_z\beta\frac{\mathfrak b_+^{Is}+\mathfrak b_-^{Is}+\mathfrak b_+^{Is}\mathfrak b_-^{Is}}{\mathfrak B_+^{Is}\mathfrak B_-^{Is}}
=2J_z\beta\Big(1-\frac{1}{\mathfrak V}\Big),\nonumber\\
&&\frac{\mathfrak b_+^{(1;2)}(x)\mathfrak b_+^{(1;-2)}(x)}{(\mathfrak B_+^{Is})^2}
+\frac{\mathfrak b_-^{(1;2)}(x)\mathfrak b_-^{(1;-2)}(x)}{(\mathfrak B_-^{Is})^2}=\frac{J_z^2\beta^2}{\mathfrak B_+^{Is}\mathfrak B_-^{Is}}
\Big(\frac{(\mathfrak b_+^{Is})^2}{\mathfrak B_+^{Is}}+\frac{(\mathfrak b_-^{Is})^2}{\mathfrak B_-^{Is}}\Big)\nonumber\\
&&=J_z^2\beta^2\Big(\frac{\mathfrak U-4}{\mathfrak V}+\frac{\mathfrak U}{\mathfrak V^2}\Big),
\end{eqnarray}
and a substitution of (67) and (68) into (57) results in
\begin{equation}
f^{(2)}(T,h)=2J_z\Big(\frac{1}{\mathfrak U}-\frac{1}{\mathfrak V}\Big)+J_z^2\beta\Big(\frac{2}{{\mathfrak U}{\mathfrak V}}-\frac{1}{\mathfrak V}\Big).
\end{equation}

According to (16), (32), (33) and (66)
\begin{equation}
\mathfrak U=2+\mathfrak T,\qquad\mathfrak V=1+{\rm e}^{-\beta J_z}+\mathfrak T,
\end{equation}
where
\begin{equation}
\mathfrak T=2{\rm e}^{-\beta J_z}\Big(\cosh{\frac{\beta h}{2}}\sqrt{\sinh^2{\frac{\beta h}{2}}+{\rm e}^{\beta J_z}}+\sinh^2{\frac{\beta h}{2}}\Big).
\end{equation}
With the use of (70) one may rewrite (69) in the form
\begin{equation}
f^{(2)}(T,h)=2J_z\mathfrak f_1+J_z^2\mathfrak f_2,
\end{equation}
where
\begin{equation}
\mathfrak f_1=\frac{{\rm e}^{-\beta J_z}-1}{(2+\mathfrak T)(1+{\rm e}^{-\beta J_z}+\mathfrak T)},\qquad
\mathfrak f_2=-\frac{\beta\mathfrak T}{(2+\mathfrak T)(1+{\rm e}^{-\beta J_z}+\mathfrak T)}.
\end{equation}
Formulas (71)-(73) express the main result of the paper.

\section{The high-temperature regime}

Rigorously speaking validity of the QTM approach was proved only for high temperatures \cite{11}. That is why a comparison between
the direct high-temperature expansion and the one which follows from (71)-(73) may be considered only as a good check of the calculations.

At high temperatures one has
\begin{equation}
f(T,h)=-\lim_{N\rightarrow\infty}\frac{1}{\beta N}\ln{{\rm tr}\Big(I-\beta\hat H+\frac{(\beta\hat H)^2}{2}\Big)}+o(\beta).
\end{equation}
Since
\begin{eqnarray}
&&{\rm tr}I=2^N, \qquad{\rm tr}\hat H=2^N\frac{N}{4}{\rm tr}H,\nonumber\\
&&{\rm tr}\hat H^2=2^N\frac{N}{4}\Big({\rm tr}H^2+{\rm tr}H_{12}H_{23}+\frac{N-3}{4}({\rm tr}H)^2\Big),
\end{eqnarray}
where the $4\times4$ matrix (here $I_n$ is $n\times n$ identity matrix and $I=I_{2^N}$)
\begin{equation}
H=\frac{J}{2}\Big({\bf S}^+\otimes{\bf S}^-+{\bf S}^-\otimes{\bf S}^+\Big)+
J_z\Big({\bf S}^z\otimes{\bf S}^z-\frac{1}{4}I_4\Big)-\frac{h}{2}\Big({\bf S}^z\otimes I_2+I_2\otimes{\bf S}^z\Big),
\end{equation}
is the Hamiltonian density related to (1) and
\begin{equation}
H_{12}=H\otimes I_2,\qquad H_{23}=I_2\otimes H.
\end{equation}

Substituting (75) into (74) and expanding the logarithm one readily gets
\begin{eqnarray}
&&f(T,h)=-\frac{\ln{2}}{\beta}-\lim_{N\rightarrow\infty}\frac{1}{\beta N}\ln\Big[1-\frac{\beta N}{4}{\rm tr}H+\frac{\beta^2N}{8}\Big({\rm tr}(H^2+H_{12}H_{23})\nonumber\\
&&+\frac{N-3}{4}({\rm tr}H)^2\Big)\Big]+o(\beta)\nonumber\\
&&=-\frac{\ln{2}}{\beta}+\frac{{\rm tr}H}{4}-\beta\Big(\frac{{\rm tr}H^2+{\rm tr}H_{12}H_{23}}{8}-\frac{3}{32}({\rm tr}H)^2\Big)\Big]+o(\beta),
\end{eqnarray}
or
\begin{equation}
f(T,h)=-\frac{2}{\beta}+\frac{{\rm tr}H}{4}+\frac{\beta}{32}\Big(3({\rm tr}H)^2-4[{\rm tr}H^2+{\rm tr}(H_{12}H_{23})]\Big)+o(\beta).
\end{equation}
At the same time from (76) follows that
\begin{equation}
{\rm tr}H=-J_z,\qquad{\rm tr}H^2=\frac{J^2+J_z^2+h^2}{2},\qquad{\rm tr}(H_{12}H_{23})=\frac{J_z^2+h^2}{2}.
\end{equation}
So a substitution of (80) into (78) yields
\begin{equation}
\tilde f(T,h)=-\frac{\beta J_z^2}{16\Delta^2}+o(\beta),
\end{equation}
and according to (47)
\begin{equation}
f^{(2)}(T,h)=-\frac{\beta J_z^2}{4}.
\end{equation}
This formula may be also readily obtained from (72), (73) and the high temperature expansion
\begin{equation}
\mathfrak T=2-J_z\beta+o(\beta),
\end{equation}
which directly follows from exact formula (71).

\section{The low-temperature polarized regime}

Before evaluating the low-temperature expansion for the free energy density of the Ising-like chain in the polarized phase (10), (11) we shall study the pure Ising chain.
A low-temperature expansion of the free energy density (4) under the condition (11) (with $J=0$) results in the formula
\begin{equation}
f^{Is}(T,h)=e_0(h)+f^{Is}_{magn}(T,h)+f^{Is}_{bound}(T,h)+f^{Is}_{scatt}(T,h)+o({\rm e}^{2\beta(J_z-h)}),
\end{equation}
where
\begin{eqnarray}
&&f^{Is}_{magn}(T,h)=-\frac{{\rm e}^{-\beta E^{Is}_{magn}}}{\beta},\qquad
f^{Is}_{bound}(T,h)=-\frac{{\rm e}^{-\beta E^{Is}_{bound}}}{\beta},\nonumber\\
&&f^{Is}_{scatt}(T,h)=\frac{3{\rm e}^{-\beta E^{Is}_{scatt}}}{2\beta}.
\end{eqnarray}
The parameters
\begin{equation}
E^{Is}_{magn}=h-J_z,\qquad E^{Is}_{bound}=2h-J_z,\qquad E^{Is}_{scatt}=2(h-J_z),
\end{equation}
are the energies of a one magnon state (a single excited spin $\dots\otimes|\uparrow\rangle\otimes|\downarrow\rangle\otimes|\uparrow\rangle\dots$),
a two-magnon bound state (two neighboring excited spins) and a two-magnon scattering states (two isolated excited spins).
The physical meaning of the expansion (84) is quite clear and expresses a subdivision of the energy spectrum on independent subsectors.
More specific is hierarchy of the terms in (84). Of course the condition (11) yields $E^{Is}_{magn}<E^{Is}_{bound},E^{Is}_{scatt}$ so that
$f^{Is}_{magn}(T,h)$ is the {\it leading} term of the expansion (if $e_0(h)$ is treated as the {\it constant} term)
\begin{equation}
f^{Is}_{lead}(T,h)=f^{Is}_{magn}(T,h).
\end{equation}
The {\it subleading} terms are however different for ferromagnets and (magnetically polarized) antiferromagnets. Namely as it follows from (85) and (86)
\begin{eqnarray}
&&f^{Is}_{sublead}(T,h)=f^{Is}_{bound}(T,h),\qquad J_z<0,\\
&&f^{Is}_{sublead}(T,h)=f^{Is}_{scatt}(T,h),\qquad h>J_z>0.
\end{eqnarray}

Let us now again turn back to the Ising-like chain. First of all let us note that reproducing the cluster expansion result of \cite{12} we have to account that the Hamiltonian (17) of \cite{12}
turns into (1) only after changing signs of the couplings $J_z\rightarrow-J_z$, $J\rightarrow-J$ and
renormalization of the Zeeman term ${\bf S}_n^z-1/2I\longrightarrow{\bf S}_n^z$. Under this procedure the low-temperature cluster expansion formula for the
free energy density of the XXZ spin chain in the polarized phase has the form (84) however with
\begin{eqnarray}
&&f_{magn}(T,h)=-\frac{{\rm e}^{\beta(J_z-h)}}{2\pi\beta}\int_{-\pi}^{\pi}{\rm e}^{-\beta J\cos{k}}dk=-\frac{{\rm e}^{\beta(J_z-h)}}{2\pi\beta}\int_{-\pi}^{\pi}{\rm e}^{\beta|J|\cos{k}}dk,\\
&&f_{bound}(T,h)=-\frac{{\rm e}^{\beta(J_z-2h)}}{2\pi\beta}\int_0^{2\pi}dk\Theta(J_z^2-J^2\cos^2{k/2}){\rm e}^{-\beta J^2/J_z\cos^2(k/2)},\\
&&f_{scatt}(T,h)=
\frac{{\rm e}^{2\beta(J_z-h)}}{4\pi\beta}\int_0^{2\pi}dk\Big({\rm e}^{-2\beta J\cos{k}}
+\frac{J_z}{\pi}\int_{-\pi}^{\pi}d\kappa\frac{{\rm e}^{2\beta J\cos{k/2}\cos{\kappa}}}
{J_z-J{\rm e}^{-i\kappa}\cos{k/2}}\Big).
\end{eqnarray}
(Here (91) is an improved version of (77) in \cite{12}). Accounting that in the Ising-like case $\Theta(J_z^2-J^2\cos^2{k/2})=1$ one readily reduce (91) to the form
\begin{equation}
f_{bound}(T,h)=-\frac{{\rm e}^{\beta(J_z-2h)}}{2\pi\beta}\int_0^{2\pi}dk{\rm e}^{-\beta J^2/J_z\cos^2(k/2)}.
\end{equation}

As in the Ising case the magnon contribution $f_{magn}(T,h)$ gives the leading low-temperature asymptotics to the free energy density while $f_{bound}(T,h)$ and $f_{scatt}(T,h)$
give a subleading one correspondingly
in ferromagnetic and polarized antiferromagnetic cases.

The expansion (84) as well as the corresponding one based on(90)-(93) are efficient only in the low-temperature regime governed by the inequality
\begin{equation}
\beta(h-J_z)\gg1,
\end{equation}
or according to (11) and (13)
\begin{equation}
k_BT\ll E_{gap}+\frac{E_{width}}{2}.
\end{equation}

In the {\it extreme} low-temperature subregime when (94) or (95) are replaced by a more strict inequality
\begin{equation}
\beta|J|\gg1\Longleftrightarrow k_BT\ll\frac{E_{width}}{2},
\end{equation}
the saddle-point integration reduces (90) to an approximative asymptotic expression
\begin{equation}
f_{magn}(T,h)\approx f^{extr}_{magn}(T,h)=-\frac{{\rm e}^{\beta(J_z+|J|-h)}}{2\pi\beta}\int_{-\infty}^{\infty}
{\rm e}^{-\beta |J|k^2/2}dk=-\frac{{\rm e}^{-\beta E_{gap}}}{\sqrt{2\pi\beta^3|J|}},
\end{equation}
which obviously can not be reproduced within the perturbation theory based on the QTM approach.

At the same time if
\begin{equation}
\frac{|J|}{h-J_z}=\frac{E_{width}}{2E_{gap}+E_{width}}\ll1,
\end{equation}
then there is additionally the {\it scaled} low-temperature subregime governed by the supplemental condition
\begin{equation}
\beta|J|\ll1\Longleftrightarrow k_BT\gg\frac{E_{width}}{2}.
\end{equation}
In this case the first two terms of the power expansion for the exponent in (90) yield
\begin{equation}
f_{magn}(T,h)\approx f^{scal}_{magn}(T,h)=-\frac{{\rm e}^{\beta(J_z-h)}}{\beta}\Big(1+\frac{\beta^2J^2}{4}\Big)
=-\frac{{\rm e}^{\beta(J_z-h)}}{\beta}\Big(1+\frac{\beta^2J_z^2}{4\Delta^2}\Big).
\end{equation}
In a similar manner (93) gives
\begin{equation}
f_{bound}(T,h)\approx f_{bound}^{scal}(T,h)=-\frac{{\rm e}^{\beta(J_z-2h)}}{\beta}\Big(1-\frac{J_z\beta}{2\Delta^2}\Big).
\end{equation}
Derivation of an analogous expansion for $f_{scatt}(T,h)$ is a bit more cumbersome. Namely
\begin{eqnarray}
&&\frac{J_z}{2\pi}\int_{-\pi}^{\pi}d\kappa\frac{{\rm e}^{-2\beta J\cos{k/2}\cos{\kappa}}}
{J_z-J{\rm e}^{-i\kappa}\cos{k/2}}=\frac{1}{2\pi}\int_{-\pi}^{\pi}d\kappa\Big(1+\frac{{\rm e}^{-i\kappa}}{\Delta}\cos{\frac{k}{2}}
+\frac{{\rm e}^{-2i\kappa}}{\Delta^2}\cos^2{\frac{k}{2}}+o\Big(\frac{1}{\Delta^2}\Big)\Big)\nonumber\\
&&\Big(1-2\beta J\cos{\frac{k}{2}}\cos{\kappa}+2\beta^2 J^2\cos^2{\frac{k}{2}}\cos^2{\kappa}+o(\beta^2 J^2)\Big)\nonumber\\
&&=1+\frac{\beta^2J_z^2-\beta J_z}{\Delta^2}\cos^2{\frac{k}{2}}+o(\beta^2J^2)+o\Big(\frac{1}{\Delta^2}\Big).
\end{eqnarray}
So according to (92) and (102)
\begin{equation}
f_{scatt}(T,h)\approx f^{scal}_{scatt}(T,h)=\frac{{\rm e}^{2\beta(J_z-h)}}{\beta}\Big(\frac{3}{2}+\frac{2\beta^2J_z^2-\beta J_z}{2\Delta^2}\Big).
\end{equation}

Let us now reproduce the asymptotic formulas (100), (101) and (103) by the QTM approach.
Representing (71) in the form
\begin{equation}
\mathfrak T={\rm e}^{-\beta J_z}({\rm e}^{\beta h}-1)\left[1+\frac{1+{\rm e}^{-\beta h}}{2}\left(\sqrt{1+\frac{{\rm e}^{\beta J_z}}{\sinh^2{(\beta h/2)}}}-1\right)\right],
\end{equation}
one readily gets at $\beta\rightarrow\infty$
\begin{equation}
\mathfrak T={\rm e}^{\beta(h-J_z)}-{\rm e}^{-\beta J_z}+1+o(\min(1,{\rm e}^{-\beta J_z})).
\end{equation}
So according to (73) and (105)
\begin{equation}
\mathfrak f_2=\beta\Big(-{\rm e}^{\beta(J_z-h)}+4{\rm e}^{2\beta(J_z-h)}\Big)+o({\rm e}^{2\beta(J_z-h)}).
\end{equation}
The corresponding expression for $\mathfrak f_1$ cardinally depends on the sign of $J_z$. Namely
\begin{equation}
\mathfrak f_1=\left\{\begin{array}{rcl}
{\rm e}^{\beta(J_z-2h)}+o({\rm e}^{\beta(J_z-2h)}),\qquad J_z<0\\
-{\rm e}^{2\beta(J_z-h)}+o({\rm e}^{2\beta(J_z-h)}),\qquad J_z>0.
\end{array}\right.
\end{equation}
Now a substitution of (106) and (107) into (72) reproduces (100), (101) and (103).

In order to visualize the difference between the low-temperature and scaled low-temperature regimes we may introduce according to (90), (97) and (100)
the following two universal functions
\begin{eqnarray}
&&g_{extr}(\zeta)=\frac{\displaystyle f_{magn}^{extr}(T,h)}{\displaystyle f_{magn}(T,h)}
=\frac{\displaystyle\sqrt{2\pi}{\rm e}^{\zeta}}{\displaystyle\sqrt{\zeta}\int_{-\pi}^{\pi}{\rm e}^{\zeta\cos{k}}dk},\nonumber\\
&&g_{scal}(\zeta)=\frac{\displaystyle f_{magn}^{scal}(T,h)}
{f_{magn}(T,h)}=\frac{\displaystyle\pi(4+\zeta^2)}{\displaystyle2\int_{-\pi}^{\pi}{\rm e}^{\zeta\cos{k}}dk}.
\end{eqnarray}
where $\zeta=\beta|J|$. According to (96), (99) and (108) the extreme low-temperature regime corresponds to $\zeta\gg1$ and $g_{extr}(\zeta)\approx1$, while the scaled low-temperature regime corresponds
to $\zeta\ll1$ and $g_{scal}(\zeta)\approx1$. Namely, as it readily follows from (108)
\begin{equation}
\lim_{\zeta\rightarrow\infty}g_{extr}(\zeta)=\lim_{\zeta\rightarrow0}g_{scal}(\zeta)=1.
\end{equation}

The plots of $g_{extr}(\zeta)$ and $g_{scal}(\zeta)$ are presented in Figure 1.
\begin{figure}
\begin{center}
\includegraphics[scale=0.55]{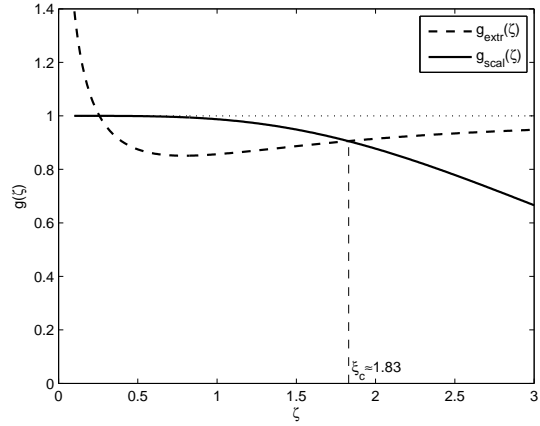}
\caption{Plots of $g_{extr}(\zeta)$ and $g_{scal}(\zeta)$.}
\end{center}
\end{figure}

From Fig.1 we see that $f^{extr}_{magn}(T,h)$ gives a better approximation to $f_{magn}(T,h)$
than $f^{scal}_{magn}(T,h)$ only for $\zeta>\zeta_c\approx1.83$.

In order to give a more detailed characterization of these approximations we suggest two relative errors functions
\begin{equation}
\varepsilon_a(\zeta)=\left|\frac{f_{magn}^a(T,h)-f_{magn}(T,h)}{f_{magn}(T,h)}\right|=|g_a(\zeta)-1|,\qquad a=extr,\,scal.
\end{equation}
Plots of these functions are presented in Figure 2 and Figure 3.

\begin{figure}
\begin{minipage}{0.5\linewidth}
\begin{center}
\includegraphics[scale=0.55]{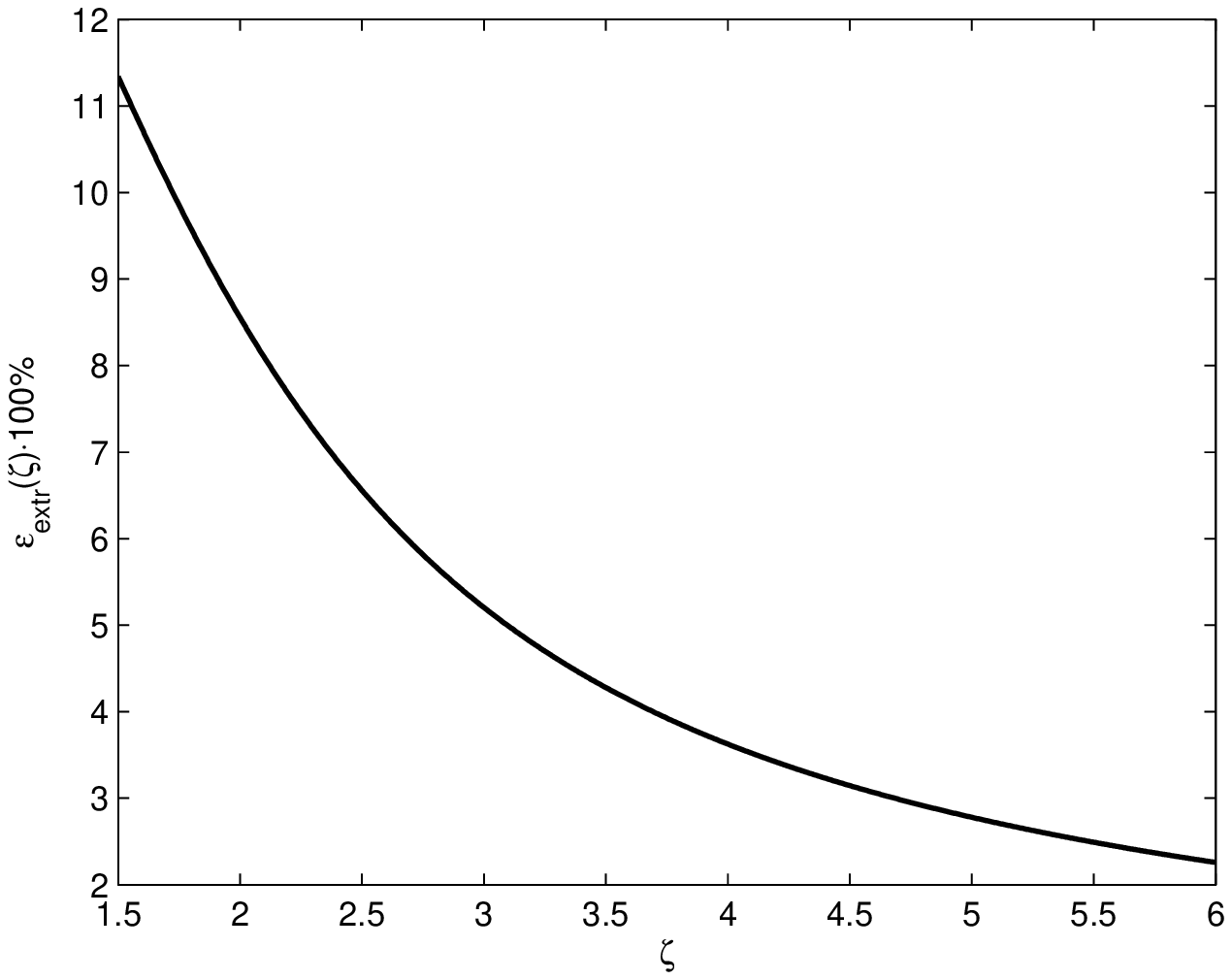}
\caption{Plot of $\varepsilon_{extr}(\zeta)\cdot100\%$.}
\end{center}
\end{minipage}
\hfil
\begin{minipage}{0.5\linewidth}
\begin{center}
\includegraphics[scale=0.55]{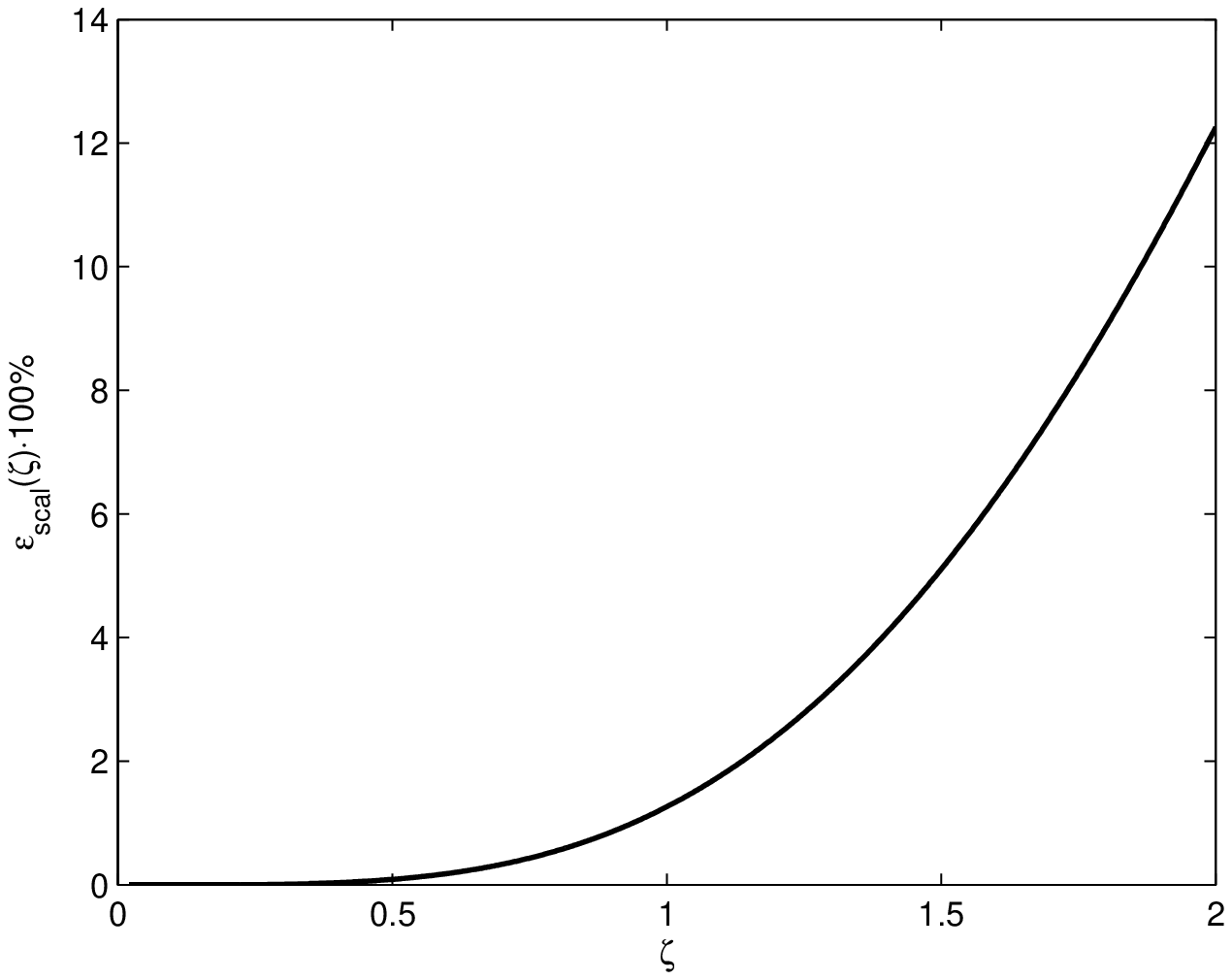}
\caption{Plot of $\varepsilon_{scal}(\zeta)\cdot100\%$.}
\end{center}
\end{minipage}
\end{figure}

\section{Summary}

In the present paper extending the result of \cite{10} belonging to the Ising chain we studied within the QTM approach a highly anisotropic Heisenberg-Ising chain
and have suggested the perturbative formula (see (71)-(73)) for the free energy density.
At high temperatures the result agrees with the direct high-temperature expansion. At low temperatures an agreement with the cluster expansion is only in the
special scaled regime (14) which may be realized only under the condition (98) and for which quantum fluctuations are small against the thermodynamical ones.
We suggest that the obtained result confirms effectiveness of the QTM approach and may be extended on perturbative evaluation of correlation functions.

\end{document}